\documentclass[aps,pre,groupedaddress,showpacs,twocolumn]{revtex4-1}
\usepackage{hyperref}
\usepackage{graphicx}
\usepackage{amsmath}
\usepackage{amssymb}

\hypersetup{colorlinks=true,citecolor=blue}

\begin{document}

\title{Autoresonant excitation of Bose-Einstein condensates}
\pacs{03.75.Kk, 05.45.-a, 05.45.Xt}
\author{S.V.~Batalov}
\email{svbatalov@imp.uran.ru}
\author{A.G.~Shagalov}
\email{shagalov@imp.uran.ru}
\affiliation{Institute of Metal Physics, Ekaterinburg 620990, Russian Federation}
\affiliation{Ural Federal University, Mira 19, Ekaterinburg 620002, Russian Federation}
\author{L.~Friedland}
\email{lazar@mail.huji.ac.il}
\affiliation{Racah Institute of Physics, Hebrew University of Jerusalem, Jerusalem 91904,
Israel}

\begin{abstract}
Controlling the state of a Bose-Einstein condensate driven by a chirped
frequency perturbation in a one-dimensional anharmonic trapping potential is
discussed. By identifying four characteristic time scales in this
chirped-driven problem, three dimensionless parameters $P_{1,2,3}$ are
defined describing the driving strength, the anharmonicity of the trapping
potential, and the strength of the particles interaction, respectively. As
the driving frequency passes the linear resonance in the problem, and
depending on the location in the $P_{1,2,3}$ parameter space, the system may
exhibit two very different evolutions, i.e. the quantum energy ladder
climbing (LC) and the classical autoresonance (AR). These regimes are
analysed both in theory and simulations with the emphasis on the effect of
the interaction parameter $P_{3}$. In particular, the transition thresholds
on the driving parameter $P_{1}$ and their width in $P_{1}$ in both the AR
and LC regimes are discussed. Different driving protocols are also
illustrated, showing efficient control of excitation and de-excitation of
the condensate.
\end{abstract}

\maketitle

\affiliation{Institute of Metal Physics, Ekaterinburg 620990, Russian
Federation}
\affiliation{Ural Federal University, Mira 19, Ekaterinburg
620002, Russian Federation}

\affiliation{Racah Institute of Physics, Hebrew University of Jerusalem,
Jerusalem 91904, Israel}


\section{\label{sec:1} Introduction}

Unique properties of Bose-Einstein condensates (BEC) attracted enormous
interest in the last decades as a very flexible framework for experimental
and theoretical research in many-body physics and a promising basis for new
technologies. Modern applications require understanding of nonlinear
dynamics of the condensates. Nonlinear dynamics is especially interesting in
anharmonic trapping potentials, when motion of the center of mass is coupled
to internal degrees of freedom \cite{Dobson_1994}, and may even become
chaotic \cite{Ott_2003}. In this paper we take advantage of the anharmonic
potential to excite a quasi-one-dimensional condensate from the ground state
to a high energy level. The basic idea is to use a driving perturbation with
a slowly varying frequency to transfer the population from the ground
quantum state to the first excited state, then to the second, and so on.
This dynamical process, when only two energy eigenstates are resonantly
coupled at a time, is the ladder climbing (LC) regime.

The classical counterpart of the ladder climbing is the autoresonance (AR),
the phenomenon discovered by Veksler and McMillan in 1944 \cite{Veksler_1944}
and referred to as the phase stability principle at the time. Nowadays, the
AR has multitude of applications in such diverse areas as hydrodynamics \cite%
{Friedland_1999,Friedland_2002,Assaf_2005}, plasmas \cite%
{Fajans_1999,Lindberg_2006,Yaakobi_2008}, magnetism \cite%
{Kalyakin_2007,Batalov_2013,Klughertz_2015}, nonlinear optics \cite%
{Barak_2009}, molecular physics \cite{Marcus_2004}, planetary dynamics \cite%
{Friedland_2001} etc. The AR in BECs was previously studied in \cite%
{Nicolin_2007} for the case of oscillating scattering length. The excitation
of a BEC from the ground state to the first energy eigenstate using optimal
control was investigated experimentally in Ref. \cite{Bucker_2013}.

In this paper we consider Gross-Pitaevskii model \cite{GP}
\begin{align}
i\hbar \Psi _{t}& +\frac{\hbar ^{2}}{2m}\Psi _{xx}-\left( U+g|\Psi
|^{2}\right) \Psi =0,  \label{eq:GP} \\
U(x,t)& =m\omega _{0}^{2}\left( \frac{x^{2}}{2}-\beta \frac{x^{4}}{4}\right)
+\varepsilon x\cos \varphi (t),  \label{eq:U}
\end{align}%
which describes a BEC in a trap with the anharmonic potential $U(x,t)$
perturbed by a small amplitude oscillating drive. Here $\beta >0$ is the
anharmonicity parameter assumed to be small. The frequency of the drive $%
\omega (t)=\dot{\varphi}(t)=\omega _{0}-\alpha t$ slowly decreases in time ($%
\alpha >0$) and passes through the linear resonance frequency $\omega _{0}$
in the problem at $t=0$. We assume that the wave function is normalized to
unity and, thus, parameter $g$ is proportional to the total number of
particles in the condensate. Both the classical AR and the quantum LC in the
linear limit ($g=0$) of Eq. \eqref{eq:GP}, i.e., the quantum Duffing
oscillator, were studied in Refs. \cite%
{Marcus_2004,Barth_2011,Andresen_2011,Shalibo_2012}.

In this work, we focus on the nonlinear effects due to the interaction of
the particles in the condensate. As a first step, we adopt the notations
used in Ref. \cite{Barth_2011} to allow comparison with the linear case. To
this end, we classify different dynamical regimes of Eq. \eqref{eq:GP} in
terms of parameters $P_{1}$, $P_{2}$ used in Ref. \cite{Barth_2011} and
introduce a new parameter $P_{3}$, characterizing the nonlinearity in the
problem. These parameters are constructed using four characteristic time
scales in the problem: the inverse Rabi frequency $T_{R}=\sqrt{2m\hbar
\omega _{0}}/\varepsilon $, the frequency sweep time scale $T_{S}=\alpha
^{-1/2}$, the anharmonic time scale $T_{A}=3\hbar \beta /(4m\alpha )$ of the
trapping potential, and the nonlinear time scale $T_{N}=g/(\hbar \alpha \ell
)$, where $\ell =\sqrt{{\hbar }/{m\omega _{0}}}$ is the characteristic width
of the harmonic oscillator. Time $T_{A}$ is the time of passage of the
frequency $\omega (t)$ through the anharmonic frequency shift between the
first two levels of the energy ladder. Similarly, $T_{N}$ is the time of
passage through the nonlinear frequency shift. Then, our three dimensionless
parameters are defined as
\begin{equation}
\begin{split}
P_{1}=\frac{T_{S}}{T_{R}}=\frac{\varepsilon }{\sqrt{2m\hbar \omega
_{0}\alpha }}& ,\quad P_{2}=\frac{T_{A}}{T_{S}}=\frac{3\hbar \beta }{4m\sqrt{%
\alpha }}, \\
P_{3}=\frac{T_{N}}{T_{S}}& =\frac{g}{\hbar \ell \sqrt{\alpha }}
\end{split}
\label{eq:par}
\end{equation}%
and characterize the strength of the drive, the anharmonicity of the
trapping potential and the nonlinearity of the condensate, respectively. We
limit our discussion to the case of the positive nonlinearity (repulsion), $%
P_{3}>0$. The scope of the paper is as follows. In Sec. \ref{sec:quantum} we
study the dynamics of our system in the energy basis of the quantum harmonic
oscillator and find a domain in $P_{1,2,3}$ parameter space where the
successive quantum energy ladder climbing process takes place. In Sec. \ref%
{sec:classic}, we discuss the opposite limit of semiclassical dynamics. The
details of our numerical simulations are presented in Sec. \ref{sec:numerics}
and the conclusions are summarized in Sec. \ref{sec:conclusions}.

\section{\label{sec:quantum}Quantum LC regime}

Here we focus on the LC regime, where the quantum nature of the system is
mostly pronounced. In this case, it is convenient to express Eq. %
\eqref{eq:GP} in the energy basis of the linear harmonic oscillator $\Psi
(x,t)=\sum_{n}c_{n}(t)\psi _{n}(x)$, yielding
\begin{equation}
\begin{split}
i\hbar \frac{dc_{n}}{dt}& =E_{n}c_{n}+\frac{\varepsilon \ell }{\sqrt{2}}%
\left( \sqrt{n+1}\,c_{n+1}+\sqrt{n}\,c_{n-1}\right) \cos \varphi \\
& +g\sum_{klm}c_{k}c_{l}c_{m}^{\ast }\int_{-\infty }^{\infty }\psi _{n}\psi
_{k}\psi _{l}\psi _{m}dx,
\end{split}
\label{eq:dcdt}
\end{equation}%
where the approximate energy levels up to linear terms in $\beta $ are given
by \cite{Landau1977quantum}
\begin{equation*}
E_{n}\approx \hbar \omega _{0}\left[ n+\frac{1}{2}-\frac{3\beta \hbar }{%
8m\omega _{0}}\left( n^{2}+n+\frac{1}{2}\right) \right] .
\end{equation*}%
Introducing new variables $B_{n}=c_{n}\,e^{i\left( \frac{E_{0}t}{\hbar }%
+n\varphi \right) }$, we rewrite Eq. \eqref{eq:dcdt} \ in the form

\begin{eqnarray}
&i&\hbar \frac{dB_{n}}{dt}=\left[ n\hbar \alpha t-\frac{3\beta \hbar }{%
8m\omega _{0}}n(n+1)\right] B_{n}  \label{eq:C1} \\
&&+\frac{\varepsilon \ell }{2\sqrt{2}}\left( \sqrt{n+1}\,B_{n+1}e^{-i\varphi
}+\sqrt{n}\,B_{n-1}e^{i\varphi }\right) \left( e^{-i\varphi }+e^{i\varphi
}\right)  \notag  \label{eq:C1} \\
&&+g\sum_{klm}B_{k}B_{l}B_{m}^{\ast }e^{i(n+m-k-l)\varphi }\int_{-\infty
}^{\infty }\psi _{n}\psi _{k}\psi _{l}\psi _{m}dx  \notag
\end{eqnarray}%
We also define the dimensionless time $\tau =\sqrt{\alpha }\,t$, coordinate $%
\xi =x/\ell $, and basis functions $\chi _{n}(\xi )=\sqrt{\ell }\,\psi
_{n}(x)$, and use the rotating wave approximation (preserve only stationary
terms in the driving and nonlinear components in Eq. \eqref{eq:C1}) to get a
dimensionless system
\begin{align}
i\frac{dB_{n}}{d\tau }& =\Gamma _{n}B_{n}+\frac{P_{1}}{2}\left( \sqrt{n+1}%
\,B_{n+1}+\sqrt{n}\,B_{n-1}\right)  \notag \\
& +P_{3}\sum_{kl}B_{k}B_{l}B_{l+k-n}^{\ast }\int_{-\infty }^{\infty }\chi
_{n}\chi _{_{k}}\chi _{_{l}}\chi _{l+k-n}d\xi ,  \label{eq:dBdt}
\end{align}%
where the frequencies $\Gamma _{n}$ are $\Gamma _{n}=n\tau -\frac{P_{2}}{2}%
\,n(n+1)$.

The resonant population transfer between levels $n-1$ and $n$ [the
Landau-Zener (LZ) transition \cite{Landau_1932}] takes place when their
time-dependent characteristic frequencies are matched: $\Gamma _{n-1}(\tau
_{n})\approx \Gamma _{n}(\tau _{n})$, i.e. in the vicinity of $\tau
_{n}=nP_{2}$. In terms of Eq. \eqref{eq:dcdt} this corresponds to the
resonance condition $E_{n}-E_{n-1}\approx \hbar \,\omega (\tau _{n})$. Note
that the anharmonicity parameter determines the time interval between
successive LZ transitions $\Delta \tau =\tau _{n}-\tau _{n-1}=P_{2}$. These
transitions can be treated independently provided their duration is much
shorter than $\Delta \tau $. As suggested in \cite{Barth_2011} for the
linear case $(P_{3}=0)$, the well-separated LZ transitions are expected when
\begin{equation}
P_{2}\gg 1+P_{1},  \label{eq:Barth}
\end{equation}%
where the right hand side is a typical duration of a LZ transition in both
adiabatic (slow passage through resonance) and fast transition limits \cite%
{Vitanov_1996}. This inequality defines the domain of essentially quantum
dynamics in the parameter space $P_{1,2}$, if $P_{3}=0$. Later in this
Section, we discuss how relation \eqref{eq:Barth} is modified in the case of
the nonlinear LZ transitions ($P_{3}\neq 0$).

Neglecting all states in \eqref{eq:dBdt}, but those with amplitudes $%
u=B_{n-1}$ and $v=B_{n}$, we obtain the nonlinear LZ-type equations
describing the isolated transition between the two states:

\begin{equation}
\begin{split}
i\frac{du}{d\tau }& =\Gamma _{n-1}\,u+P_{3}\left(
|u|^{2}I_{n-1}+2|v|^{2}J_{n}\right) u+\frac{P_{1}\sqrt{n}}{2}v, \\
i\frac{dv}{d\tau }& =\Gamma _{n}\,v+P_{3}\left(
|v|^{2}I_{n}+2|u|^{2}J_{n}\right) v+\frac{P_{1}\sqrt{n}}{2}u,
\end{split}
\label{eq:NLZ}
\end{equation}%
where $J_{n}=\int |\chi _{n-1}|^{2}|\chi _{n}|^{2}d\xi $, $I_{n}=\int |\chi
_{n}|^{4}d\xi $, 
The nonlinear LZ model attracted significant attention recently \cite%
{Liu_2002,Zhang_2008,Trimborn_2010}, especially in the context of BECs in
optical lattices. Nonlinearity may change the dynamics of the system
significantly compared to the linear case. One deviation from the linear
case is the breakdown of adiabaticity due to the bifurcation of nonlinear
stationary states \cite{Wu_2000,Zobay_2000}. In this paper we focus on
another property of the nonlinear LZ model, namely, the AR.

Equations \eqref{eq:NLZ} can be simplified using the conservation of the
total probability {$K_{n}=|u|^{2}+|v|^{2}$}. Introducing the fractional
population imbalance $S=({|v|^{2}-|u|^{2}})/{K_{n}}$ and the phase mismatch $%
\delta =\arg \left( {v/u}\right) $, we obtain a set of real equations
\begin{equation}
\begin{split}
\frac{dS}{d\tau }& =-\mu _{n}\sqrt{1-S^{2}}\sin \delta , \\
\frac{d\delta }{d\tau }& =\frac{P_{3}K_{n}f_{n}}{2}S-(\tau -\tau _{\ast })+%
\frac{P_{1}\sqrt{n}\,S}{\sqrt{1-S^{2}}}\cos \delta ,
\end{split}
\label{eq:res}
\end{equation}%
where $f_{n}=4J_{n}-I_{n-1}-I_{n}$ and $\tau _{\ast
}=nP_{2}-P_{3}K_{n}(I_{n}-I_{n-1})/2$ . The AR in a similar system was
studied in \cite{Barak_2009}. Its main characteristic is a continuing phase
locking as the phase mismatch $\delta $ remains bounded due to a persistent
self-adjustment of the nonlinear frequency of the driven system to slowly
varying driving frequency, i.e., $P_{3}K_{n}f_{n}S/2\approx (\tau -\tau
_{\ast })$. However, the slow passage through resonance does not guarantee
the AR. Indeed, if one assumes that only state $u$ is populated initially,
that is $S\rightarrow -1$, then the AR requires the driving parameter $P_{1%
\text{ }}$to exceed a certain threshold \cite{Barak_2009}
\begin{equation}
P_{1}>P_{1,cr}=\frac{0.82}{\sqrt{nP_{3}f_{n}K_{n}}}.  \label{eq:P1cr}
\end{equation}%
This condition \eqref{eq:P1cr} shows that the AR is an essentially nonlinear
phenomenon and disappears when $P_{3}\rightarrow 0$.

A typical dynamics of the nonlinear LZ model is illustrated in Fig. \ref%
{fig:add}.
\begin{figure}[tbh]
\includegraphics[scale=0.35]{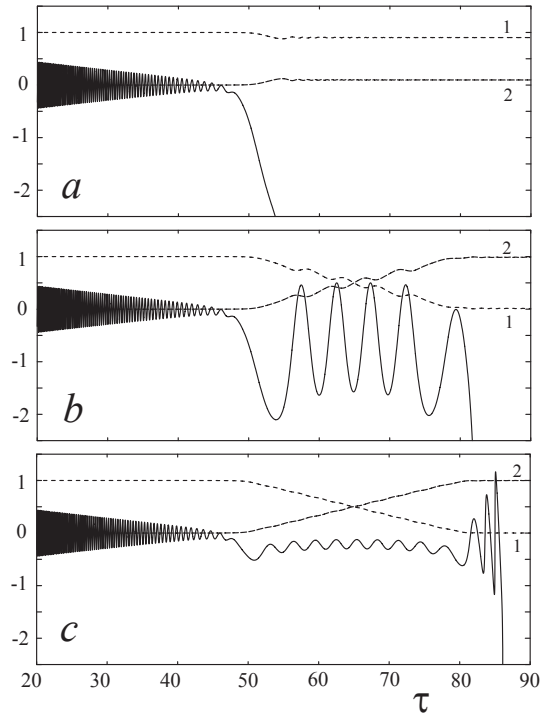}
\caption{The phase mismatch $\protect\delta $ (solid lines) and amplitudes
(1 -- $|u|^{2}$ and 2 -- $|v|^{2}$, dashed lines) versus slow time $\protect%
\tau$ in the nonlinear Landau-Zener transition for $P_{2}=50$, $P_{3}=300$.
Panel (a) below the AR threshold $P_{1}=0.14<P_{1,cr}=0.15$, panel (b) just
above the threshold $P_{1}=0.16$, and panel (c) well above the threshold $%
P_{1}=0.31$ when the AR phase-locking is observed; the parameters are $n=1$,
$K_{1}=1$, and $f_{1}\approx 0.1$.}
\label{fig:add}
\end{figure}
For a given nonlinearity $P_{3}$, the threshold condition (\ref{eq:P1cr})
separates two different types of evolution of the system. If $P_{1}<P_{1,cr}$
(see Fig. \ref{fig:add}$a$), the passage through the resonance at $\tau
\approx \tau _{\ast }-P_{3}K_{n}f_{n}/2\approx 50$ yields a small excitation
and a fast growth of the phase mismatch $\delta $ between the two states
(see solid line in Fig. \ref{fig:add}$a$). In contrast, we observe
synchronization and nearly complete transition between the states when the
coupling parameter exceeds the threshold, $P_{1}>P_{1,cr}$. Figures \ref%
{fig:add}$b$ and \ref{fig:add}$c$ illustrate this effect just beyond the
threshold and far from the threshold, respectively. One can see that above
the threshold the phase mismatch is bounded, exhibiting the phase-locking
phenomenon characteristic of the AR. One can also see that the amplitudes
vary significantly during the transition from $n-1$ to $n$ state. The
phase-locking is destroyed only after the system almost completely transfers
to the new state. As long as the phase-locking is sustained, the population
imbalance increases on average as $S(\tau )\approx {2(\tau -\tau _{\ast })}/{%
P_{3}f_{n}K_{n}}$ with some superimposed modulations. In particular, $%
|u|^{2}\approx |v|^{2}$ at time $\tau \approx \tau _{\ast }\ $($\tau _{\ast
}=65$ in our examples). Since the maximal change of the population imbalance
is $\Delta S=2$, the duration of the complete autoresonant transition can be
estimated as
\begin{equation*}
\tau _{AR}=P_{3}f_{n}K_{n}.
\end{equation*}

\begin{figure}[tbh]
\includegraphics[scale=0.6]{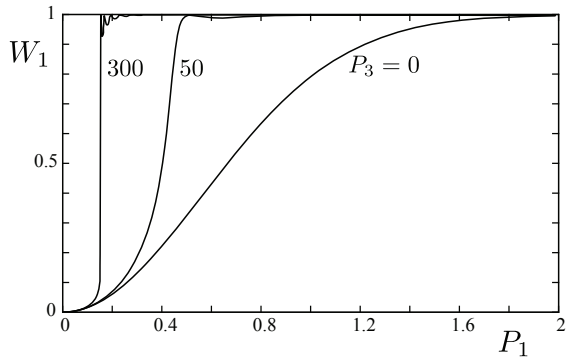}
\caption{The probability $W_{1}(P_{1})$ of the $0\rightarrow 1$ transition
for different values of the nonlinearity parameter $P_{3}=0,50,300$ from
numerical simulations of Eq. \eqref{eq:NLZ} for $n=1$, $K_{1}=1$.}
\label{fig:prob}
\end{figure}

The most important characteristic of the LZ model is the transition
probability $W_{n}={|v(\infty )|^{2}}/{K_{n}}$ for finding the system in the
upper state if it was in the lower state initially. In the linear limit,
this probability is given by the famous LZ formula
\begin{equation}
W_{n}=1-e^{-\frac{n\pi P_{1}^{2}}{2}}.  \label{eq:Wlinear}
\end{equation}%
The numerical integration of the nonlinear LZ model \eqref{eq:NLZ} gives the
transition probability shown in Fig.~\ref{fig:prob}. The curve corresponding
to $P_{3}=0$ coincides with the linear LZ result \eqref{eq:Wlinear}. The
transition probability steepens as the nonlinearity $P_{3}$ increases and
tends to a step-like function in the strongly nonlinear limit. The front of
this step corresponds to the onset of the AR at $P_{1}\approx P_{1,cr}$.
Once phase-locked, the system remains in this state until almost complete
population inversion is achieved, as indicated by transition probability
(the height of the step) close to unity in Fig.~\ref{fig:prob}. This means
that the threshold $\eqref{eq:P1cr}$ obtained in Ref. \cite{Barak_2009} in a
small-amplitude limit, i.e. assuming $S\approx -1+\delta S$, $\delta S\ll 1$%
, is applicable to the fully nonlinear equations \eqref{eq:res} as well.

For interpreting numerical simulations covering both the linear and
nonlinear LZ transitions we redefine the threshold $P_{1,cr}$ as the $P_{1}$
value corresponding to $50\%$ transition probability, i.e., $%
W_{n}(P_{1,cr})=1/2$. Using this definition we numerically solve Eq.%
\eqref{eq:NLZ} to find the threshold values $P_{1,cr}$ and compare these
results with the theoretical prediction \eqref{eq:P1cr} in Fig.~\ref%
{fig:threshold}. One can see that in the strongly nonlinear limit $%
P_{3}f_{n}\gg 1$ the numerical results reproduce Eq. \eqref{eq:P1cr}. On the
other hand, for small nonlinearity the threshold approaches a constant $%
P_{1,cr}\approx \sqrt{2\ln {2}/\pi n}$ corresponding to the linear LZ
formula \eqref{eq:Wlinear}.

\begin{figure}[tbh]
\includegraphics[scale=0.6]{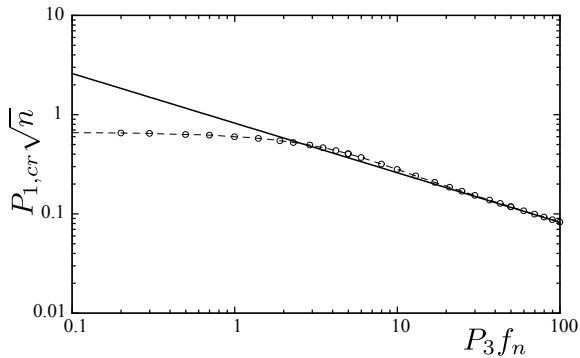}
\caption{The rescaled critical drive parameter versus rescaled nonlinearity
parameter. Solid line -- theory, Eq. \eqref{eq:P1cr}; circles -- numerical
simulations of Eq. \eqref{eq:NLZ}.}
\label{fig:threshold}
\end{figure}

The original system \eqref{eq:dcdt} also allows excitation from the ground
state to a $N$-th energy state via a sequence of $N$ independent LZ
transitions with probability
\begin{equation}
W=\prod_{n=1}^{N}W_{n}(P_{1},P_{3}).  \label{eq:W}
\end{equation}%
The equation $W(P_{1},P_{3})=1/2$ defines the threshold $P_{1,LC}(P_{3})$
for this LC process. As it was found numerically \cite{Barth_2011} in the
linear $P_{3}=0$ case, the product \eqref{eq:W} quickly converges for $%
N\geqslant 5$ and one finds $P_{1,LC}\approx 0.79$. On the other hand, in
the strongly nonlinear limit one can approximate the transition probability
by the Heaviside step function $W_{n}(P_{1},P_{3})\approx
H(P_{1}-P_{1,cr}(P_{3},n))$, where $P_{1,cr}$ is given by Eq. \eqref{eq:P1cr}%
. Since $P_{1,cr}$ decreases with the transition number $n$ and every
transition leads to nearly complete population inversion, the product %
\eqref{eq:W} can be replaced by the single $n=1$ term $W\approx W_{1}$ and
the capture into the LC regime occurs after the first transition. In this
approximation the threshold is simplified
\begin{equation*}
P_{1,LC}\approx P_{1,cr}(P_{3},n=1)=\frac{0.82}{\sqrt{P_{3}f_{1}K_{1}}}.
\end{equation*}%
Note that the threshold width, defined as the inverse slope of the
transition probability at $W=1/2$, equals $\Delta P_{1,LC}=0.66$ in the
linear case \cite{Barth_2011} and tends to zero in the strongly nonlinear
regime.

Similarly to the linear case, we can assume that the successive transitions
in the nonlinear regime will be well-separated provided the time between two
successive transitions satisfies $\Delta \tau \gg \tau _{AR}$, i.e., $%
P_{2}\gg f_{1}K_{1}P_{3}$. This estimate can be simplified because $f_{1}=%
\sqrt{2/\pi }/8\approx 0.1$ and we can set $K_{1}=1$. Combining the above
inequality with the linear result \eqref{eq:Barth} we find the condition for
essentially quantum dynamics in the $P_{1,2,3}$ parameter space:
\begin{equation}
P_{2}\gg 1+P_{1}+0.1P_{3}.  \label{eq:qcl}
\end{equation}%
This inequality together with the condition $P_{1}>P_{1,LC}$ defines the
region of the parameter space, where efficient excitation of quantum states
in the model \eqref{eq:GP} via the autoresonant LC is achieved.

\section{\label{sec:classic}Semiclassical regime}

If the anharmonicity parameter $P_{2}$ of the trap decreases, the two-level
approximation employed in the previous section breaks down as several levels
can resonate with the drive simultaneously. In the limit $P_{2}\ll 1$, the
number of coupled levels is so large that the dynamics becomes
semiclassical. The linear $P_{3}=0$ case in this problem was already studied
in Ref. \cite{Barth_2011}. It was shown that the autoresonant excitation of
BEC oscillations is possible provided the drive strength $P_{1}$ exceeds the
classical autoresonance threshold for the Duffing oscillator \cite%
{Fajans_2001}
\begin{equation}
P_{1,AR}=\frac{0.82}{\sqrt{P_{2}}}.  \label{eq:P1AR}
\end{equation}%
In this regime, the center of mass of the condensate oscillates in the trap
with an increasing amplitude. The frequency of these oscillations remains
close to the driving frequency during the whole excitation process, despite
the variation of the driving frequency. In this section we discuss the
threshold value $P_{1,AR}$ in the nonlinear case $P_{3}>0$.

Consider the Wigner representation of our problem \cite{hiley2004phase}:
\begin{equation}
\frac{\partial f(x,p,t)}{\partial t}=\{H,f\}_{MB},  \label{eq:dfdt}
\end{equation}%
where $f(x,p,t)$ is the Wigner function, $H$ is the Hamiltonian
\begin{equation*}
H=-\frac{\hbar ^{2}}{2m}\frac{\partial ^{2}}{\partial x^{2}}+U(x,t)+g|\Psi
(x,t)|^{2}
\end{equation*}%
and $\{H,f\}_{MB}$ denotes the Moyal bracket. Since the Moyal bracket
reduces to the Poisson bracket in the semiclassical limit $\hbar \rightarrow
0$: $\{H,f\}_{MB}\approx \{H,f\}+O(\hbar ^{2})$, equation \eqref{eq:dfdt}
reduces to the Liouville equation
\begin{equation}
\frac{\partial f}{\partial t}+\frac{p}{m}\frac{\partial f}{\partial x}-\frac{%
\partial }{\partial x}\left( U+g|\Psi |^{2}\right) \frac{\partial f}{%
\partial p}\approx 0,  \label{eq:dfdt_2}
\end{equation}%
where in addition to the external potential $U(x,t),$ we have the
self-potential $V=g|\Psi |^{2}$. This equation is reminiscent of the Vlasov
equation for an ensemble of particles in the combined external and
self-potentials. Note that the self-potential can be expressed via the
Wigner function
\begin{equation*}
|\Psi (x,t)|^{2}=\int_{-\infty }^{\infty }f(x,p,t)dp,
\end{equation*}%
transforming \eqref{eq:dfdt_2} into a closed integro-differential form.

The characteristics (classical trajectories) for Eq.~\eqref{eq:dfdt_2} are
given by
\begin{equation}
\frac{d^{2}x}{dt^{2}}+\frac{1}{m}\frac{\partial U}{\partial x}+\frac{g}{m}%
\frac{\partial |\Psi |^{2}}{\partial x}=0.  \label{eq:A}
\end{equation}%
Suppose one starts in a localized state, so that the Wigner function has a
local maximum at the phase-space point $[x_{0},p_{0}]$. In the semiclassical
limit, the Wigner function is expected to continue having a local maximum
near the phase-space point $[x(t),p(t)]$ moving along the classical
trajectory starting at $[x_{0},p_{0}]$. Near this point,%
\begin{equation*}
|\Psi (x+s,t)|^{2}\approx const-\kappa s^{2}/2
\end{equation*}%
and, thus, the term 
$(g/m)\partial |\Psi |^{2}/\partial x$ in Eq. \eqref{eq:A} vanishes along
the trajectory of the maximum of $f$. Consequently, the nonlinearity
(characterized by parameter $P_{3}$) in the semiclassical regime does not
affect the evolution of the maximum. Then it also does not change the
threshold of the AR \eqref{eq:P1AR} and should not shift the transition
probability versus $P_{1}$ in contrast to the quantum regime (see Fig.~\ref%
{fig:prob}). We confirm these conclusions in numerical simulations in the
following section.

\section{\label{sec:numerics} Numerical simulations}

In this section we present numerical simulations of the original
Gross-Pitaevskii equation (\ref{eq:GP}) in our driven problem. We rewrite
this equation in a dimensionless form using the same $\xi =x/\ell $, but a
different dimensionless time $T=\omega _{0}t$:%
\begin{align}
i\Phi _{T}& +\frac{1}{2}\Phi _{\xi \xi }-\left( \tilde{U}+Q_{3}|\Phi
|^{2}\right) \Phi =0,  \label{eq:GPnd} \\
\tilde{U}(\xi ,T)& =\frac{\xi ^{2}}{2}-Q_{2}\frac{\xi ^{4}}{4}+Q_{1}\xi \cos
\tilde{\varphi}(T),  \notag
\end{align}%
where $\Phi =\sqrt{\ell }\;\Psi $, $\tilde{\alpha}=\alpha /\omega _{0}^{2}$,
$Q_{1}=\sqrt{2\tilde{\alpha}}\,P_{1}$, $Q_{2}=\frac{4}{3}\sqrt{\tilde{\alpha}%
}\,P_{2}$, $Q_{3}=\sqrt{\tilde{\alpha}}\,P_{3}$ and $\tilde{\varphi}(T)=T-{%
\tilde{\alpha}\,T^{2}}/2$. The simulations are based on the standard
pseudo-spectral method \cite{canuto_spectral} with explicit 4-th order
Runge-Kutta algorithm and adaptive step size control. The ground state of
the harmonic oscillator was used as the initial condition:
\begin{equation}
\Phi (\xi ,0)=\pi ^{-1/4}\;e^{-\xi ^{2}/2}.  \label{incond}
\end{equation}%
The state of the condensate was analyzed by calculating its energy
\begin{equation*}
E=\int_{-\infty }^{\infty }\left( \frac{1}{2}|\Phi _{\xi }|^{2}+\tilde{U}%
|\Phi |^{2}+\frac{Q_{3}}{2}|\Phi |^{4}\right) d\xi ,
\end{equation*}%
the amplitudes $c_{n}(t)$ in the basis of Hermite functions, and the Wigner
distribution.

\begin{figure}[tbh]
\includegraphics[scale=0.80]{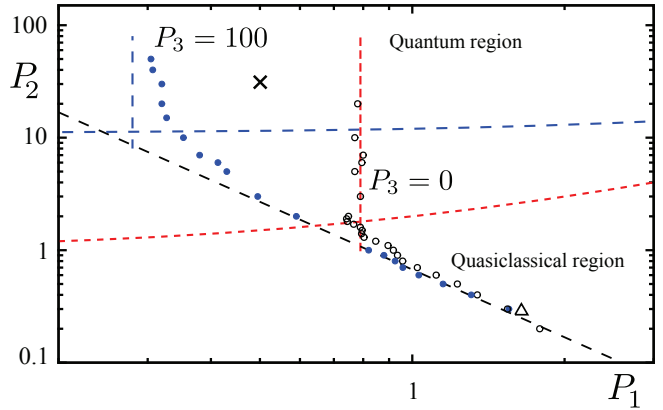}
\caption{Different domains of phase-locking transition in the driven
Gross-Pitaevskii equation \eqref{eq:GPnd} in the linear ($P_{3}=0$, black
open circles) and nonlinear ($P_{3}=100$, filled blue circles) cases. The
symbols $\times $ and $\triangle $ show parameters used in subsequent
figures \protect\ref{fig:control} and \protect\ref{fig:control2}. }
\label{fig:regimes}
\end{figure}

In order to study various regimes of excitation of a condensate, we
performed a series of numerical simulations by varying parameters $P_{1}$
and $P_{2}$ in the linear ($P_{3}=0$) and nonlinear cases. The results of
the simulations are presented in Fig.~\ref{fig:regimes}. The circles in the
figure correspond to parameters yielding $50\%$ probability of capture into
either the classical AR or the quantum LC regime.
The classical autoresonance threshold \eqref{eq:P1AR} is shown by the black
long-dashed line. The two roughly horizontal lines $P_{2}=1+P_{1}+0.1\,P_{3}$
separate the regions of the quantum and semiclassical dynamics in the linear
($P_{3}=0$, red line) and nonlinear ($P_{3}=100$, blue line) regimes. The
vertical lines show the theoretical LC transition thresholds $P_{1,LC}$ for
the linear and nonlinear regimes.

One can see that the classical AR threshold \eqref{eq:P1AR} yields a good
approximation for the transition boundary for $P_{2}$ versus $P_{1}$ in both
the linear and the nonlinear cases. In the case $P_{3}=100$, the nonlinear
ladder climbing transitions emerge at significantly smaller values of the
driving parameter $P_{1}$ and larger anharmonicity $P_{2}$, compared to the
linear case. This is in agreement with the shift of the threshold in the
nonlinear model of the autoresonant LZ transitions discussed above (see Fig.
2).

\begin{figure}[tbh]
\includegraphics[scale=0.45]{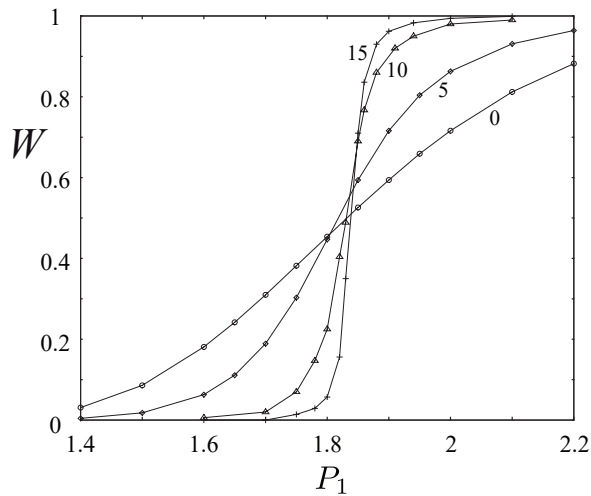}
\caption{The transition probability versus the driving parameter $P_{1}$ for
$P_{2}=0.2$ and $P_{3}=0,5,10,15$ from numerical integration of Eq.~
\eqref{eq:GPnd}. }
\label{fig:W}
\end{figure}

The important change due to the nonlinearity in the problem is the decrease
of the width of the transition region. This effect is illustrated in Fig.~%
\ref{fig:W} for the case of the semiclassical autoresonant transition. The
width of the transition for the nonlinear case decreases rapidly with the
increase of the nonlinearity parameter $P_{3}$ and the transition
probability assumes a nearly step-like shape for $P_{3}>15$. One also
observes that the threshold location, where the probability crosses $50\%$,
only slightly changes with the variation of $P_{3}$ as discussed at the end
of Sec. II.

The effect of the narrowing of the transition width in the semiclassical
regime with the increase of the nonlinearity parameter can be associated
with the improved stability of the autoresonant classical trajectories
described by Eq. (\ref{eq:A}) . Indeed, in dimensionless variables, Eq. (\ref%
{eq:A}) for the characteristics of Eq. (\ref{eq:dfdt_2}) becomes
\begin{equation}
\xi _{TT}+\xi -Q_{2}\xi ^{3}+Q_{3}\frac{\partial |\Phi (\xi ,T)|^{2}}{%
\partial \xi }+Q_{1}\cos \widetilde{\varphi }(T)=0.  \label{eq:char}
\end{equation}%
Let the trajectory of the autoresonant maximum of the Wigner function be $%
\xi _{0}(T)$. The dynamics of this trajectory is described by%
\begin{equation}
\xi _{0TT}+\xi _{0}-Q_{2}\xi _{0}^{3}+Q_{1}\cos \widetilde{\varphi }(T)=0
\label{artr}
\end{equation}%
subject to $\xi _{0}=\xi _{0T}=0$ at large negative $T$ and is not affected
by the nonlinearity, as described above. For studying the the evolution of a
deviation $\eta =\xi -\xi _{0}(T)$ from $\xi _{0}(T)$, we linearize Eq. (\ref%
{eq:char}) around $\xi _{0}$ and assume $|\Phi (\xi ,T)|^{2}\approx
const-\kappa \eta ^{2}/2$ near the maximum, to get
\begin{equation}
\eta _{TT}+[1-3Q_{2}\xi _{0}^{2}-\kappa Q_{3}]\eta =0.  \label{eta}
\end{equation}%
We analyze the solutions of system (\ref{artr}), (\ref{eta}) in the
Appendix. Numerically, in the vicinity of the threshold for $P_{3}=0$, we
observe the development of instability of $\eta $. By writing $\xi
_{0}=a\cos\theta$, Eq. (\ref{eta}) assumes the form of the
Mathieu-type equation with slowly varying parameters%
\begin{equation}
\eta _{TT}+\left\{ 1-\frac{3Q_{2}a^{2}}{2}-\kappa Q_{3}-\frac{3}{2}%
Q_{2}a^{2}\cos [2\theta (T)]\right\} \eta =0.  \label{eq:paramet}
\end{equation}%
We show in the Appendix that this equation predicts parametric-type
instability for $P_{3}=0$ and we attribute the existence of the width in the
transition to autoresonance as seen in Fig.~\ref{fig:W} to this effect. Note
that the addition of the nonlinearity (the term $-\kappa Q_{3}$ in the last
equation) shifts the eigenfrequency so that the parametric resonance can be
avoided and characteristic trajectories with nearby initial conditions
remain close to the classical autoresonant trajectory, narrowing the
transition width as seen in Fig. \ref{fig:W}. For the parameters in this
figure, the system stabilizes at $P_{3}>0.3$ (see the Appendix).
Nonetheless, the deviation from the autoresonant state is still large until $%
P_{3}>10$, when the autoresonant transition width practically disappears.

\begin{figure}[tbh]
\includegraphics[scale=0.65]{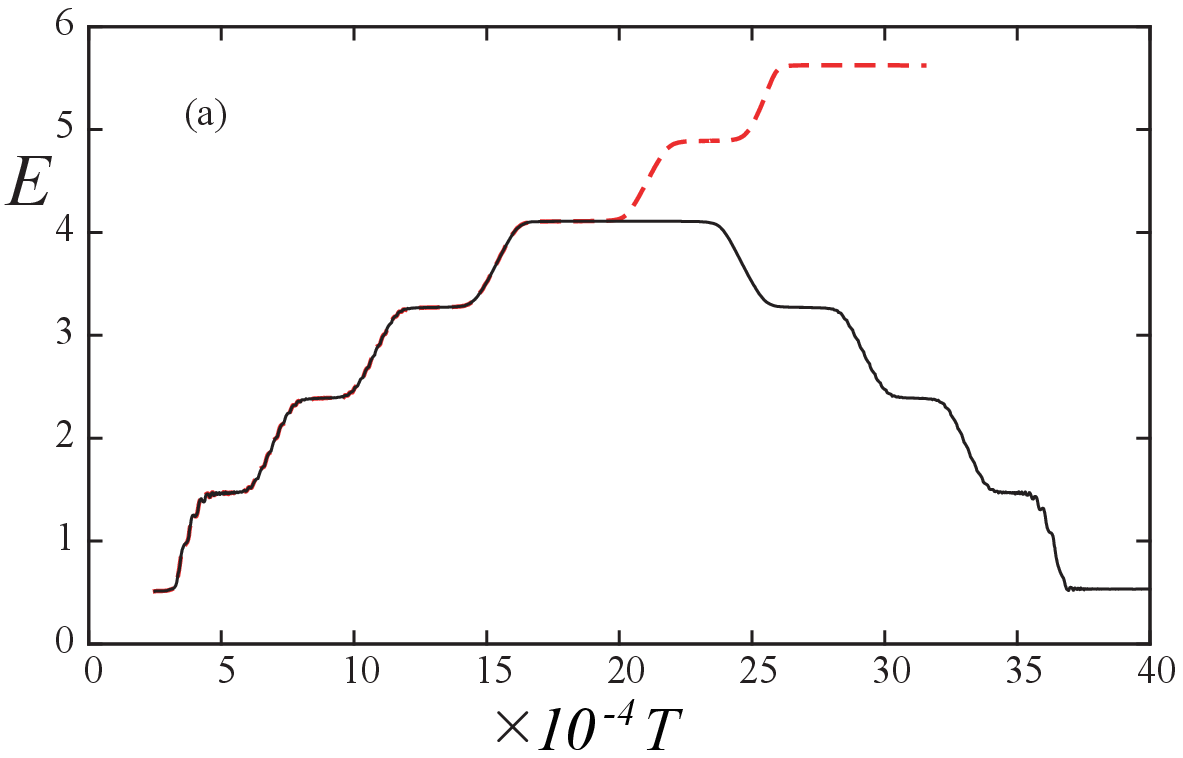} %
\includegraphics[scale=0.45]{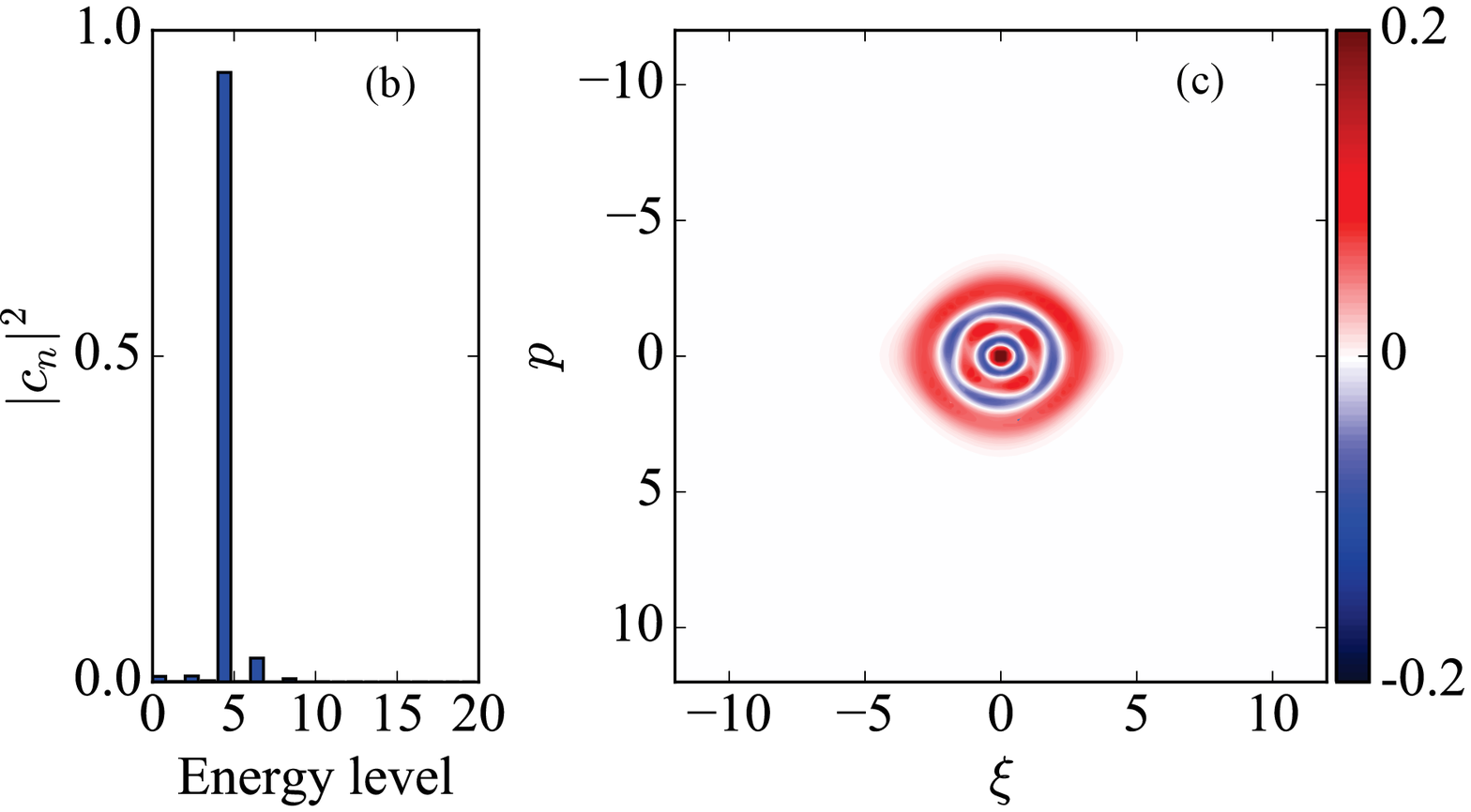}
\caption{Numerical solutions of Eq.~\eqref{eq:GPnd} in the LC regime for two
scenarios. In the first scenario [dashed red lines in panel (a)], constant
driving frequency chirp rate $\tilde{\protect\alpha}=10^{-6}$ and $P_{1}=0.5$%
, $P_{2}=30$, $P_{3}=100$ (as shown by cross symbol in Fig.~\protect\ref%
{fig:regimes}) are used. In the second scenario [solid line in panel (a)] we
set $\tilde{\protect\alpha}=0$ at $T=1.8\times 10^{5}$ and keep it constant
until $T=2.2\times 10^{5}$. We restore the original value of $\tilde{\protect%
\alpha}$, but with the opposite sign at $T=2.2\times 10^{5}$. The
distribution of the populations of the energy levels and the Wigner
quasi-probability distribution at $T=2\times 10^{5}$ for the second scenario
are shown in panels (b) and (c), respectively. Here $T=\protect\omega _{0}t=%
\protect\tau /\protect\sqrt{\tilde{\protect\alpha}}$.}
\label{fig:control}
\end{figure}

Our next numerical simulation shows that the system can be fully controlled
by the AR in both the quantum and semiclassical regimes. Two protocols of
such a control of the LC dynamics are shown in Fig. \ref{fig:control}.
Parameters $P_{i}$ are chosen so that both condition \eqref{eq:qcl} and $%
P_{1}>P_{1,LC}$ are satisfied. Due to a relatively strong anharmonicity, the
nonlinear LZ transitions are well separated in time. One can see that the
energy in the system grows step-by-step from one energy level to another. In
the first protocol, the linear frequency variation with a constant chirp
rate $\alpha $ [red dashed line in Fig. \ref{fig:control}(a)] results in the
excitation of the condensate to the sixth energy level. In the second
protocol (solid line) we first excite the system to the fourth energy state
by decreasing driving frequency until $T=1.8\times 10^{5}$, then keep the
driving frequency constant for the time span of $\Delta T=4\times 10^{4}$,
and finally return the system to the ground state by increasing the driving
frequency back to its original value. One can see that the quantum state of %
\eqref{eq:GP} can be efficiently controlled as long as the LC autoresonant
conditions are met and the frequency and phase of the driving are
continuous. The quantum state in the first protocol at $T=2\times 10^{5}$ is
further illustrated in Figs.~\ref{fig:control} (b) and (c) showing the
distribution of populations of different levels and the Wigner function.
Small deviations of the solution from the fourth eigenfunction $\chi
_{4}(\xi )$ of the linear harmonic oscillator are due to the anharmonicity
and the nonlinearity in the problem.

\begin{figure}[tbh]
\includegraphics[scale=0.65]{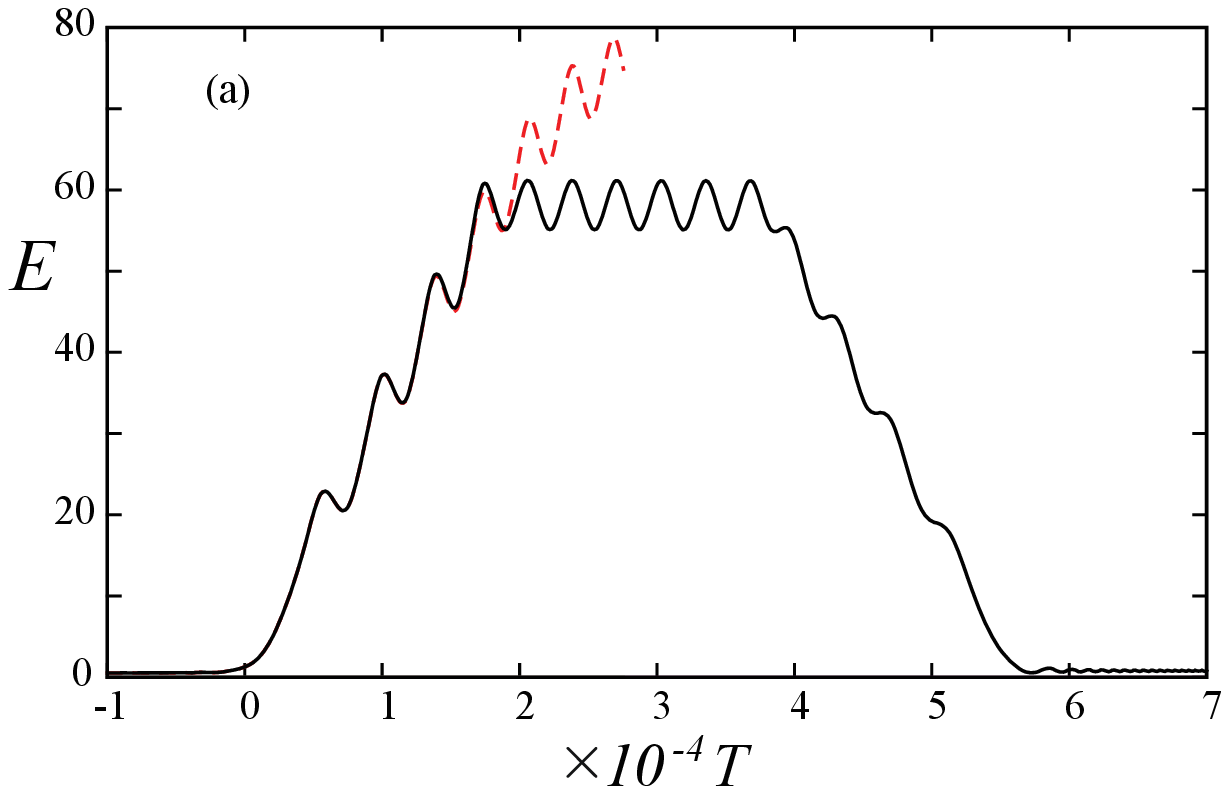} %
\includegraphics[scale=0.4]{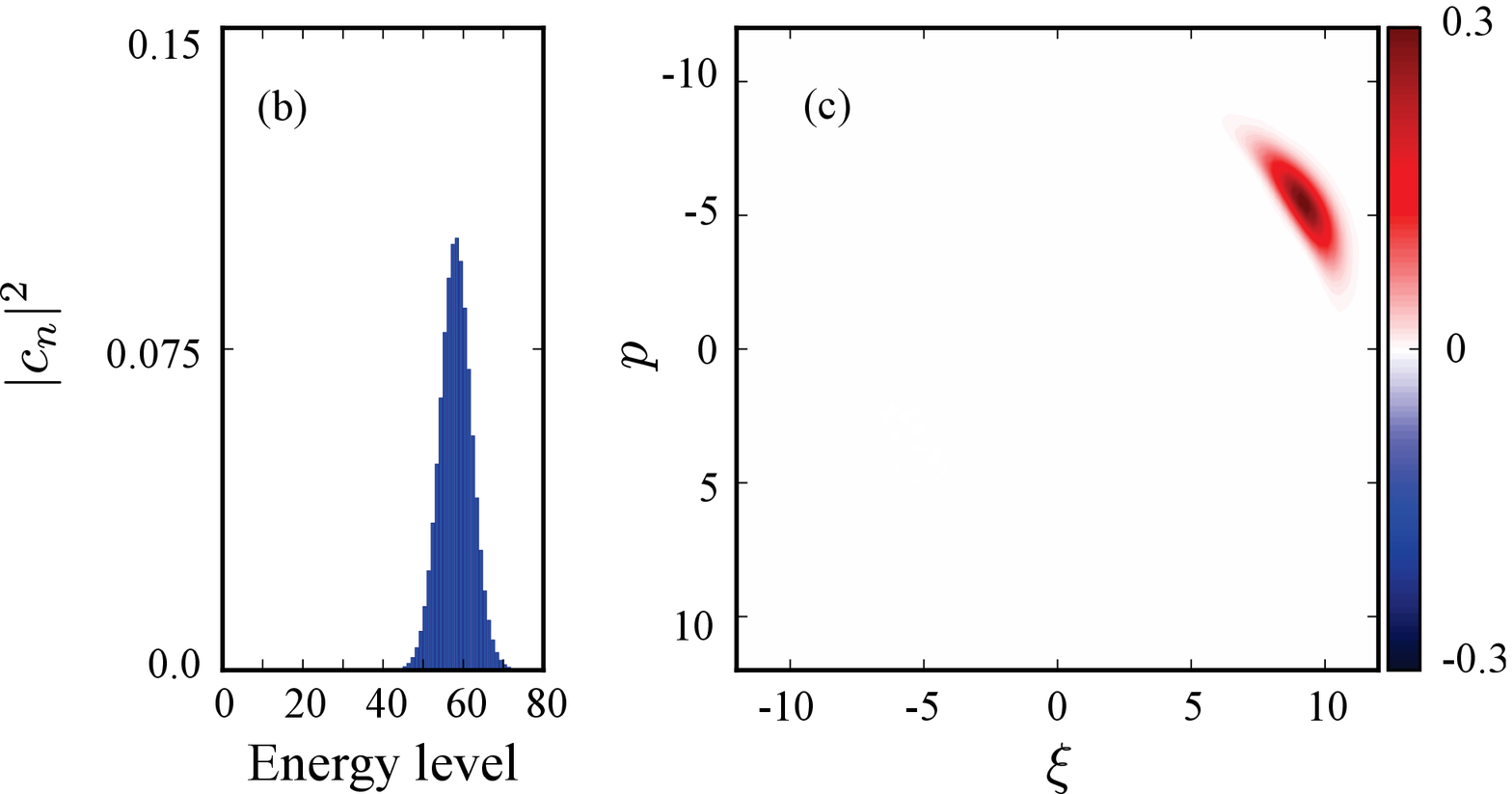}
\caption{The numerical solutions of Eq.~\eqref{eq:GPnd} in the semiclassical
regime for parameters $\tilde\protect\alpha=10^{-6}$, $P_1=1.6$, $P_2=0.3$, $%
P_3=100$ (represented by the triangle in Fig.~\protect\ref{fig:regimes}). In
the second scenario (solid line) the frequency chirp is set to zero from $%
T=1.8\times 10^{4}$, but restored to the original value with opposite sign
at $T=3.8\times 10^{4}$. The distribution of populations of energy levels
and the Wigner function at $T=2.8\times 10^4$ for the second scenario are
shown in panels (b) and (c), respectively.}
\label{fig:control2}
\end{figure}

A similar control protocols for the semiclassical case are illustrated in
Fig.~\ref{fig:control2}. One can see that resulting wave function has a
Gaussian (Poissonian in the early stages of dynamics) population
distribution, characteristic of coherent states. The Wigner function is
positive everywhere (on the computational grid) and close to the $n$%
-squeezed coherent state. The second protocol (solid line in Fig.~\ref%
{fig:control2}) demonstrates that a more complex control scenarios are
possible in the semiclassical case, as long as the driving parameter is
within the region of autoresonant dynamics and the driving phase and
frequency are continuous functions of time.

\section{\label{sec:conclusions}Conclusions}

In conclusion, we have studied the effect of the particles interaction on
the excitation of Bose-Einstein condensate in a anharmonic trap under
chirped-frequency perturbation. We have identified three dimensionless
parameters $P_{1,2,3}$ [see Eqs. (\ref{eq:par})] characterizing the driving
strength, the anharmonicity and the strength of the interaction to show that
there exist two very different regimes of excitation in this parameters
space, i.e., the quantum-mechanical ladder climbing (LC) and the
semiclassical autoresonance (AR). The transition boundary to the
semiclassical AR in the $P_{1,2}$ parameter space is independent of the
nonlinearity parameter $P_{3}$. In contrast, the LC transition boundary is
significantly affected by the strength of the interaction of the particles,
because the underlying nonlinear Landau-Zener (LZ) transitions behave
differently than their linear counterpart. In the limit of strong
interaction, the nonlinear LZ transition probability as a function of the
driving strength parameter $P_{1}$ approaches the Heaviside step function
due to the nonlinear phase-locking. We have also found that in both the
quantum and the semiclassical regimes the width\ $\Delta P_{1}$ of the
transition decreases as the strength of the interaction increases. In the
quantum limit this effect is related to the autoresonance of the nonlinear
LZ transitions, while in the semiclassical limit, the effect is due to the
wave packet stability enhancement by avoiding parametric resonance between
the center-of-mass motion and the internal dynamics of the condensate.

Possible applications of the results of this paper may include a control of
the quantum state of BECs and the implementation of precision detectors
based on either the LC or the AR. Unlike the noninteracting case \cite%
{Murch_2010}, the resolution of such a detector is not limited by quantum
fluctuations if the particles interaction is strong enough.

\section*{ACKNOWLEDGEMENTS}

This work was supported by the Russian state assignment of FASO
No.01201463332 and by the Israel Science Foundation Grant No. 30/14.

\section*{APPENDIX: Stability of autoresonant trajectories.}

\begin{figure}[tbh]
\includegraphics[scale=0.7]{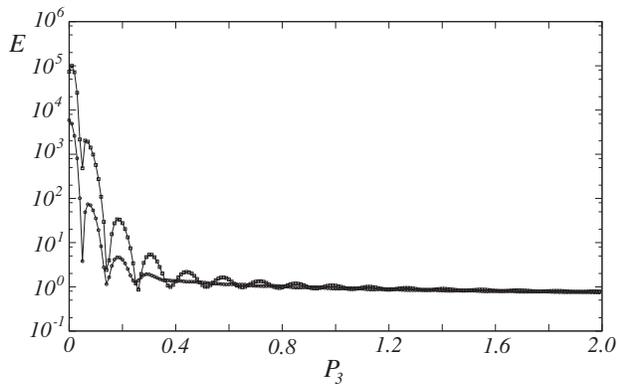}
\caption{The energy $E=\frac{1}{2}\left\langle \protect\eta _{T}^{2}+\protect%
\eta \right\rangle $ averaged over the integration time versus $P_{3}$ from
the numerical solution of Eqs. \eqref{eq:char},\eqref{eq:paramet} near the
threshold of the autoresonance. The lines with squares and circles
correspond to $\protect\alpha =10^{-4}$ and $\protect\alpha =10^{-6}$,
respectively.}
\label{fig:appendix}
\end{figure}

\bigskip Here we discuss the Mathieu-type Eq. (\ref{eq:paramet})
\begin{equation}
\eta _{TT}+[B-C\cos (2\theta )]\eta =0,
\label{A1}
\end{equation}%
where $B=1-\frac{3Q_{2}a^{2}}{2}-\kappa Q_{3}$, $C=\frac{3}{2}Q_{2}a^{2}$,
describing the deviation\ $\eta $ of the trajectory from the autoresonant
solution $\xi_{0}=a\;\cos \theta$ given by Eq. (\ref{artr}).


For sufficiently small amplitudes $a$ this solution is described by \cite%
{Scholarpedia}
\begin{eqnarray}
\frac{{da}}{dT} &=&(Q_{1}/2)\sin \Phi ,  \label{A3} \\
\frac{{d\Phi }}{dT} &=&\alpha T-(3Q_{2}/8)a^{2}+(Q_{1}/2a)\cos \Phi ,
\label{A4}
\end{eqnarray}%
where $\Phi =\theta -\widetilde{\varphi }+\pi $ is the phase mismatch, which
starts at zero initially (at large negative $t$, where $a$ is also zero). We
are interested in passage through resonance when parameter $Q_{1}$ is close
to the autoresonance threshold $Q_{1th}$ (from either above or below). In
both of these cases, \ $\Phi $ slowly increases initially and then passes $%
\pi /2$. Then when slightly above the threshold, $\Phi $ starts oscillating,
but remains bounded ($|\Phi |<\pi $), while below the threshold $\Phi $
continues to increase reaching $\pi $ and the oscillator fully dephases from
the drive.

At this point we observe that there are three parameters ($\alpha
,Q_{1},Q_{2}$) in system (\ref{A3}), (\ref{A4}). However, by introducing the
slow time $\tau ={\alpha ^{1/2}T}$ and rescaled amplitude $A=\sqrt{3/8}$ $%
\alpha ^{-1/4}Q_{2}^{1/2}a$ we obtain a single parameter system

\begin{eqnarray}
\frac{{dA}}{d\tau } &=&\mu \sin \Phi ,  \label{A5} \\
\frac{{d\Phi }}{d\tau } &=&\tau -A^{2}+(\mu /A)\cos \Phi ,  \label{A6}
\end{eqnarray}%
where
\begin{equation}
\mu =\sqrt{\frac{3}{32}}Q_{1}Q_{2}^{1/2}{\alpha ^{-3/4}.}  \label{ttt}
\end{equation}%
The autoresonant threshold in this system is $\mu _{th}=0.41$ \cite%
{Scholarpedia}, which upon the return to the original parameters $P_{1,2}$
yields (\ref{eq:P1AR}). Note that at the threshold ($\mu =\mu _{th}$) the
rescaled problem has no free parameters.

Next we discuss Eq. (\ref{A1}). Here, using (\ref{A4}), we have
\begin{equation}
\frac{{d\theta }}{dT}=\frac{{d}\widetilde{\varphi }}{{dT}}+\frac{d\Phi }{dT}%
=1-\alpha T+\frac{d\Phi }{dT}=1-\frac{3}{8}Q_{2}a^{2}+\frac{Q_{1}}{2a}\cos
\Phi  \label{A7}
\end{equation}%
and assume that $S=-\frac{3}{8}Q_{2}a^{2}+\frac{Q_{1}}{2a}\cos \Phi $ is
small compared to unity in the region of dephasing. This suggests to
transform from $T$ to $\theta $ in (\ref{A1}) and approximate the problem by
the Mathieu equation with slow coefficients

\begin{equation}
\frac{{d^{2}\eta }}{{d\theta ^{2}}}+[B^{\prime }-C^{\prime }\cos (2\theta
)]\eta \approx 0  \label{A8}
\end{equation}%
where to lowest order in small parameters
\begin{eqnarray}
B^{\prime } &=&\frac{B}{(1+S)^{2}}\approx 1-\frac{3}{4}Q_{2}a^{2}-\frac{Q_{1}%
}{a}\cos \Phi -\kappa Q_{3}  \label{A9} \\
C^{\prime } &=&\frac{C}{(1+S)^{2}}\approx C=\frac{3}{2}Q_{2}a^{2}
\label{A10}
\end{eqnarray}%
From the theory of the Mathieu equation (with fixed parameters, which we
assume locally in our case), the stability condition of the solution of (\ref%
{A8}) is \cite{Hayashi}
\begin{equation}
B^{\prime }<1-C^{\prime }/2  \label{A12}
\end{equation}%
or%
\begin{equation}
\kappa Q_{3}>\max \left( -\frac{Q_{1}}{a}\cos \Phi \right) .  \label{A13}
\end{equation}%
Finally, in the last equation we use the rescaled amplitude $A$ instead of $%
a $ to get the dimensionless condition

\begin{equation}
\kappa P_{3}>P_{1}P_{2}^{1/2}\max \left( -\frac{\cos \Phi }{A}\right) _{\mu =\mu
_{th}}  \label{A11}
\end{equation}%
where the rescaled parameters are $P_{1}=Q_{1}/(2{\alpha )^{1/2}}$, $%
P_{2}=(3/4)Q_{2}/{\alpha ^{1/2}}$, $P_{3}=Q_{3}/{\alpha ^{1/2}}$ and the
value of $\max \left( -\frac{\cos \Phi }{A}\right) _{\mu =\mu _{th}}=0.37$
is found numerically by solving (\ref{A5}), (\ref{A6}). Thus, for $\kappa=0$, the
solution remains stable only if $\Phi <\pi /2$ and, since we approach this
value in the vicinity of the threshold as mentioned above, one encounters
instability near the threshold, explaining the appearance of the width of
the autoresonant transition as some initial conditions dephase. On the other
hand, a sufficiently large $\kappa P_{3}$ stabilizes the solution and the width of
the autoresonance threshold disappears. For the parameters of Fig. \ref%
{fig:W} one finds that the solution is stable for $P_{3}>0.27$. Finally we
check these conclusions by solving the set (\ref{artr}) and (\ref{eta})
numerically. The results of these simulations are presented in Fig. 8,
showing the energy $E=\frac{1}{2}\left\langle \eta _{T}^{2}+\eta
\right\rangle $ averaged over the integration time versus $P_{3}$ for the
parameters of Fig. \ref{fig:W} and initial conditions $\eta =1$, $\eta
_{T}=0 $ and integrating between $T=-2\times 10^4$ and $4\times 10^4$ for $\alpha =10^{-4%
\text{ }}$and $T=-2\times 10^5$ and $4\times 10^5$ for $\alpha =10^{-6}$. One can see the
transition to instability at $P_{3}<0.3$.

\end{document}